\documentclass[12pt]{article}

\usepackage{epsf,epsfig}

\newcommand{\autor}[1] {\begin{center}{\bf \lineskip .3cm #1} \end{center}}
\newcommand{\address}[1] {\begin{center}  {\normalsize \bf \it #1 }\end{center}}
\def\simgt{\rlap{\lower 3.5 pt \hbox{$\mathchar \sim$}} \raise 1pt \hbox {$>$}}
\def\simlt{\rlap{\lower 3.5 pt \hbox{$\mathchar \sim$}} \raise 1pt \hbox {$<$}}

\def\3half{\textstyle\frac32}
\begin{document}
\begin{titlepage}
\vspace{0.7cm}
\begin{center}
\large\bf\boldmath Importance of Higher Twist Effects to Understand
Charmed Color-Suppressed $B$ Decays \unboldmath
\end{center}
\vspace{0.8cm}
\autor{{Zuo-Hong Li,$^{1,\,2,}$\footnote{Email:
lizh@ytu.edu.cn}~Zhong-Qian Su$^{1}$ and Jian-Ying
Cui$^{1,}$}\footnote{Email: cjy@ytu.edu.cn}} \vspace{0.7cm}
\address{$^1$ Department of Physics, Yantai University,
Yantai 264005,China\,\footnote{Mailing address}}
\address{$^2$ CCAST (World Laboratory), P.O.Box 8730, Beijing 100080,
China}
\vspace{1.0cm}
\begin{abstract}
\vspace{0.2cm}
Working within the framework of the QCD light cone sum rules (LCSR),
we compute and discuss the nonfactorizable higher twist effect in
$\bar{B}^0\to D^{*0}\pi^0$ to make an all-around examination of its
role in the charmed color-suppressed $B$ decays. Analogously to the
case of $\bar{B}^0\to D^{0}\pi^0$, such effect turns out to be of
the same strong phase as the factorizable amplitude, and modifies
constructively the magnitude by order $(40-90)\%$ so that the
effective coefficient $a_2^{f}=C_1+C_2/3$ receives a positive
correction comparable numerically with it. Nonleading as the soft
effect in question is, our findings for it, along with the previous
LCSR analyses of $\bar{B}^0\to D^{0}\pi^0$, are suggestive of the
dominance of soft exchanges in these charmed color suppressed B
decays. Also, the emphases are put on importance of understanding
intensively various related higher twist and transverse momentum
effects to interpret the data on $B \to
D^{0(*0)}(\pi^0,~\eta,~\eta'),~ J/\psi~ K^{(*)}.$

\end{abstract}
\vfil
\end{titlepage}
\newcommand{\vspu}{\vspace*{5mm}}
The formulation of the QCD factorization (QCDF) \cite{QCDF} and
soft-collinear effective theory (SCET)~\cite{SCET} has greatly
renewed and deepened our understanding of QCD dynamics in the
two-body hadronic decays of B-mesons, and as such considerably
enhanced our confidence in precisely estimating the decay amplitudes
and CP-asymmetry effects. Especially, ones have essentially changed
minds in understanding of the charmed color-suppressed B decays and
realized that the nonfactorizable soft exchange contributions
dominate over the hard-exchange ones. For all these progresses,
however, being unable to deal with non-perturbative QCD dynamics
from the first principle makes yet it difficult to theoretically
interpret the experimental observations of the color suppressed
process $\bar{B}^0\to D^{0(*0)}\pi^0$ \cite{data}. One attempts to
make sense of them by a global analysis of the data on $B\to
D^{(*)}\pi$ decays, either in a model-independent manner \cite{Xing,
Neubert, HY1}, or in the framework of SCET \cite{SCET1} and
Perturbative QCD \cite{PQCD}. The leading soft effects in such
decays are supposed to include final-state rescattering processes
and nonfactorizable contributions induced by the color-octet
operators. Resorting to various phenomenological approaches, such as
one-particle exchange model \cite{HY2} and Regge theory
\cite{Regge}, one models the soft contributions from the former. QCD
sum rule approach is initially used \cite{Haperin} and QCD
light-cone sum rule (LRSR) method \cite{LCSR} is subsequently
applied \cite{Kh1} to obtain an estimate of the latter. Focusing on
the $\bar{B}^0\to D^0\pi^0$ case, the author of \cite{Haperin} finds
that the soft exchanges between the color-octet quark pair $c\bar q
$ and the pion, which concern higher-twist components of the pion,
provide a negative correction at a level of $70\%$ of the
factorizable amplitude, while the LCSR approach predicts a positive
numerical effect at $(50-110)\%$ order \cite{Cui}. As compared to
the $B\to J/\Psi K$ case where the effect in question contributes to
the magnitude by $(30-70)\%$ \cite{Melic}, the higher-twist effect
in $\bar{B}^0\to D^0\pi^0$ is, as it were, more important, as
expected. Also, the similar calculations are carried out for $B\to
\pi\pi$ \cite{Kh1, Kh2} and $B\to K\pi$ \cite{Huang}, but the
resulting numerical impacts are found to be far less than those in
both $\bar{B}^0\to D^0\pi^0$ and $B\to J/\Psi K$ cases, a result in
accordance with the argument for QCD factorization.

Whereas the nonfactorizable higher-twist effect, as estimated, plays
a role comparable numerically with the factorizable amplitudes of
the color-suppressed $\bar{B}^0\to D^0\pi^0$ and $B\to J/\Psi K$
decays, in the letter we would like to examine how it affect
$\bar{B}^0\to D^{*0}\pi^0$, to get an all-around understanding of
its role in the charmed color suppressed $B$ decays

We start with looking briefly back at the dynamical features of the
color suppressed $\bar{B}^0\rightarrow D^{0(*0)}\pi^{0}$. In the
Fierz transformed form, the relevant effective weak Hamiltonian
reads \cite{Hamiltonian}
\begin{eqnarray}
{\cal H}_{\rm W} = {G_F\over\sqrt{2}}\, V_{cb}V_{ud}^*
\Big[(C_1(\mu)+\frac{1}{3}C_2(\mu))O_1(\mu)+2 C_2(\mu)\tilde{{\cal
O}_1}(\mu)\Big]\;, \label{eq:hamilton}
\end{eqnarray}
where $C_{1,2}$ are the Wilson coefficients, $V_{ij}$ the CKM matrix
elements, and
\begin{eqnarray}
{\cal O}_1= (\bar c\Gamma^\mu u)(\bar d\Gamma_\mu
b),~~~~~~~~~\tilde{{\cal O}_1}= (\bar c\frac{\lambda_a}{2}\Gamma_\mu
u)(\bar d\frac{\lambda_a}{2}\Gamma^\mu b),
\end{eqnarray}
the color-single and -octet quark operators, respectively, with
$\Gamma_\mu=\gamma_\mu(1- \gamma_5)$ and $\lambda_a$ being the color
$SU(3)$ matrices. It is clear that the transitions induced by
$\tilde{{\cal O}_1}$ come from gluon exchanges and thus the
corresponding matrix elements are non-factorizable. Since the
ejected $D^{0(0*)}$ in the decays is a heavy meson inside which the
two quarks are asymmetric in momentum distribution, the
non-factorizable vertex diagrams prove themselves to be QCD infrared
divergent so that QCD factorization breaks down \cite{QF1}. This
hints at the dominance of soft dynamics. If we introduce an
effective coefficient $a_2$ such that the decay amplitude of
$\bar{B}^0\to D^{*0}\pi^{0}$ is made be of the following
parametrization:
\begin{equation}
{\cal A}(\bar{B}^{0}(p+q)\to D^{*0}(\epsilon,p)\pi^0(q))=-G_F
V_{cb}V^*_{ud}\,\epsilon\cdot q\, m_{D^*} f_{D^*}\,F_1^{B\to
\pi}(m_{D^*})a_2,
\end{equation}
with $\epsilon$, $m_{D^*}$ and $f_{D^*}$ being the polarization
vector, mass and decay constant of $ D^*$ meson, respectively, and
$F_1^{B\to \pi}(m_{D^*}^2)$ the $B\to \pi$ form factor defined as
\begin{equation}
\langle\pi(q)\mid\bar{u}\gamma_{\mu}b\mid
B(p+q)\rangle=(2q+p)_{\mu}F_1^{B\to
\pi}(p^2)+p_{\mu}\frac{m_B^2-m_{\pi}^2}{p^2}(F_0^{B\to
\pi}(p^2)-F_1^{B\to \pi}(p^2)),
\end{equation}
$a_2$ is actually supposed to absorb all the nonfactorizable
contributions including higher twist component we are to calculate.

Let's go over to LCSR calculation of the nonfactorizable
higher-twist effect involved in $\bar{B}^0\to D^{*0}\pi^{0}$. With
this end in view, we opt for the following correlator:
\begin{eqnarray}
\Pi_{\mu}(p,q,k)&=&i^2\int\!\!\int d^4 x d^4 y
e^{-i(p+q)x+i(p-k)y}\nonumber\\
 &&\times\langle \pi^0(q) | T
\{\bar{u}(y)\gamma_{\mu} c(y), {\tilde{O}_1}(0),m_b \bar{b}(x)i
\gamma_5 d(x)\} | 0 \rangle,\label{eq:cor1}
\end{eqnarray}
where the current operators $\bar{u} \gamma_{\mu}c$ and $\bar{b} i
\gamma_5 d$ are used to interpolate $D^{*0}$ and $\bar{B}^{0}$ meson
fields respectively, and an unphysical momentum $k$ is introduced to
make B channel free of pollution by $D^*\pi$ resonances, but it must
disappear from the physical $\bar{B}^0\to D^{*0}\pi^{0}$ amplitude
in the dispersion integral. Schematically, the higher twist
contributions to the correlator can be illustrated by Feynman
diagrams in Fig.1. The correlation function $\Pi_{\mu}$ is a
functions of 6 independent invariants, which, adapting to the
present purpose, are to be chosen as $ p^2, q^2,(p+q)^2, (p-k)^2,
k^2$ and $P^2=(p+q-k)^2$. At small light-cone distances $x^2\sim
y^2\sim (x-y)^2\sim 0$, or equivalently in large and space-like
regions of $P^2,(p+q)^2$ and $(p-k)^2$, the Operator Product
Expansion (OPE) is applicable to Eq.(\ref{eq:cor1}). Furthermore,
not violating the validity of OPE we simply let $k^2$ be zero, and
take chiral limit approximation $q^2=0$ to the pion.

Inserting on the right hand side of (\ref{eq:cor1}) a complete set
of hadronic states with $D^{*0}$ quantum numbers, we obtain the
phenomenological expression $\Pi_{\mu}^{H(D^*)}$,
\begin{eqnarray}
\Pi_{\mu}^{H(D^*)}((p-k)^2,(p+q)^2,P^2,p^2)&= &\frac{
~m_{D^{*0}}f_{D^{*0}}}
{m^2_{D^*}-(p-k)^2}~G_{\mu}((p+q)^2,P^2,p^2)\nonumber\\
&+&\int\limits_{s_0^{H(D^*)}}^{\infty}
\!\!\!ds~\frac{\rho^{H(D^*)}_{\mu}(s,(p+q)^2,P^2,p^2)}{s-(p-k)^2}\,,\label{eq:hadron}
\end{eqnarray}
where
\begin{eqnarray}
G_{\mu} &=& i\sum_{\lambda}\epsilon^{\lambda}_{\mu}\int d^4x\;
e^{-i(p+q)x} \langle D^{*0}(\epsilon^{\lambda}, p-k)~\pi^0(q)|T\{
\tilde{O}_1(0),~m_b
\bar{b}(x) i \gamma_5d(x)\} | 0 \rangle\nonumber\\
&=&\sum_{\lambda}\epsilon^{\lambda}_{\mu}~\Pi((p+q)^2,P^2,p^2,\epsilon^{*(\lambda)}
\cdot Q_i) \label{eq: TP}
\end{eqnarray}
is a two point function with $Q_i=p,~q$ and~$k$,~
$\rho^{H(D^*)}_{\mu}$~ and $s_0^{H(D^*)}$ are respectively the
hadronic spectral function and threshold mass squared of the higher
states in $D^{*}$ channel. The integral of (\ref{eq:hadron}),
invoking quark-hadron duality, is supposed to coincide with the
$s\geq s_0^{D^*}$ part of the QCD spectral representation of
$\Pi_{\mu}$
\begin{equation}
\frac{1}{\pi}\int\limits_{s_0^{D^*}}^{\infty} \!\!\!ds~\frac{{\rm
Im}_s\Pi^{(QCD)}_{\mu}(s,(p+q)^2,P^2,p^2)}{s-(p-k)^2},\nonumber
\label{eq:du1}
\end{equation}
with $s_0^{D^*}$ being the QCD effective threshold in $D^{*}$
channel, and $\frac{1}{\pi}{\rm Im}_s \Pi^{(QCD)}_{\mu}$ the related
QCD spectral density. Then we equate (\ref{eq:hadron}) with the QCD
answer for (\ref{eq:cor1}) and make the Borel transformation
$(p-k)^2\rightarrow T$, yielding
\begin{equation}
G_{\mu}(p+q)^2,P^2,p^2)= \frac{1}{\pi
m_{D^*}f_{D^*}}\int_{m_c^2}^{s_0^{D^*}}ds~ e^{(m_{D^*}^2-s)/T}{\rm
Im}_{s} \Pi_{\mu}^{(QCD)}(s,(p+q)^2,P^2,p^2). \label{eq:rPI}
\end{equation}

To proceed, saturating the two-point function $G_{\mu}$ with the
intermediate states of $\bar{B}^0$ quantum numbers, we have its
hadronic expression in which the matrix element $\langle
D^{*0}(\epsilon^{\lambda},p-k)\pi^0(q)|\tilde{O}_1
|\bar{B}^0(p+q)\rangle$ enters the lowest pole term. When the
condition $(p+q)^2=(p+q-k)^2=m_B^2$ is imposed on the
$\bar{B}^0(p+q)\to D^{*0}(\epsilon^{\lambda},p-k)\pi^0(q)$
transition, the unphysical momentum $k$ vanishes automatically so
that the corresponding matrix element is made physical. On some
algebraic manipulations we see easily that the required physical
amplitude can be parameterized in term of a function factor
$M^{(p-k)}$ in front of $(p-k)_{\mu}$ in the kinematic decomposition
of $\sum\epsilon_{\mu}^{\lambda}\langle D^{*0}(\epsilon^{\lambda},
p-k)\pi^0(q)|\tilde{O}_1 |\bar{B}^0(p+q)\rangle$. This denotes that
we need only to calculate the related part $\Pi^{(p-k)}_{QCD}$ of
$\Pi^{(QCD)}_{\mu}$. Furthermore, to apply quark hadron duality to
$B$ channel we stick down ${\rm Im}_{s}
\Pi_{\mu}^{(QCD)}(s,(p+q)^2,P^2,p^2)$ in a form of dispersion
integral in variable $(p+q)^2$, and make the analytic continuation
$P^2\longrightarrow m_B^2$, keeping $(p+q)^2$ fixed and letting
$p^2=m_{D^*}^2$. Then using the standard procedure the desired sum
rule can be achieved, if the relevant QCD double spectral density
$\rho_{QCD}(s,s',m_B^2,p^2)$ is known, where $s'$ is a spectral
variable in $B$ channel.

Now we embark on the computation of $\Pi^{(p-k)}_{QCD}$ to extract
$\rho_{QCD}$. It is advisable that we work with the fixed-point
gauge in which the light cone expansion of quark propagator
\cite{Propagator}, which takes in a correction term due to the
interaction with one background-field gluon. After some lengthy
calculations we obtain the twist-3 contribution:
\begin{eqnarray}
\Pi^{(p-k)}_{{\rm tw3}} &=& - \frac{m_b f_{3 \pi}}{8\sqrt{2} \pi^2}
\int_{m_c^2}^{\infty} \frac{ds}{s - (p-k)^2}
\int_{\frac{m_c^2}{s}}^{1} dt\int_{\chi(s,t,m_B^2)}^1
du\int_{\chi(s,t,m_B^2)}^u dv\frac{1}{[m_b^2 - (p+q
u)^2]v^2}\nonumber\\
&\times &\phi_{3\pi}(1 -u, u-v, v )\left\{(s - \frac{m_c^2}{t})(1-t)
+ 2t[(p+q)^2 - p^2]\right . \nonumber\\
&\times &\left . ( 2 v - \chi(s,t,m_B^2))\right\}
\end{eqnarray}
In the above equation, $f_{3\pi}$ is a nonperturbative quantity
defined by the vacuum-pion matrix element of the local operator
$\bar{u} \sigma_{\mu \nu} \gamma_5 G_{\alpha \beta}d$, $\phi_{3\pi}$
refers to a pionic twist-3 distribution amplitude related to the
nonlocal operator $\bar{d}(0) \sigma_{\mu \nu} \gamma_5 G_{\alpha
\beta}(v y)u(x)$ \cite{LCF}, and
$\chi=(s-\frac{m_c^2}{t})/(s-m_B^2)$. The fact that the threshold
$s^{D}_0$ is far below the squared mass of b quark allows us to make
an expansion of (10) in $\chi$. Then passing over those terms
vanishing after Borel transformations we have, up to order ${\cal
O}(\chi^3)$,
\begin{eqnarray}
\hspace*{0.3cm} \rho_{tw3}(s,s',m_B^2,p^2)
&=& \frac{m_b f_{3 \pi}}{16\sqrt{2}\pi^2(s'-p^2)}
\int_{\frac{m_c^2}{s}}^{1}
 dt                                      
\nonumber \\
 & &\hspace*{-3cm} \times \Bigg\{
 2\int_0^u \frac{dv}{v^2} \phi_{3\pi}(1 -u, u-v, v)
 \left [(1-t)\left(s -  \frac{m_c^2}{t}\right) + 2 v t(s' - p^2) \right ]\nonumber \\
 & & \hspace*{-3cm}
 +2\left [(1-3t)(s'-p^2)\int_0^u \frac{dv}{v^2} \phi_{3\pi}(1 -u, u-v, v)
\right . \nonumber \\
 & & \hspace*{-3cm}  \left . + (t-1)\left ( s - \frac{m_c^2}{t} \right ) \left ( \frac{1}{v^2} \phi_{3\pi}(1 -u, u-v, v) \right )_{v=0}
 \right ] \chi(s,t,m_B^2)\nonumber \\
 & &  \hspace*{-3cm} +\Bigg[(t-1)\left ( s - \frac{m_c^2}{t} \right )
 \frac{\partial}{\partial v}\left(\frac{1}{v^2} \phi_{3\pi}(1 -u, u-v, v)
 \right)\nonumber\\
 & &  \hspace*{-3cm} +2(2t-1)(s'- p^2) \frac{1}{v^2}\phi_{3\pi}(1 -u, u-v,
 v)\Bigg]_{v=0}\chi^2\Bigg\}.
\label{eq:tw3exp}
\end{eqnarray}
with $u=(m_b^2-p^2)/(s'-p^2)$. Following the same line the twist-4
contribution is attainable. Here we do not give it any more to save
some space.

At the end, the required nonfactorizable effect in $ \bar{B}^0\to
D^{*0}\pi^0 $ can be numerically estimated by the following LCSR
answer:
\begin{eqnarray}
\hspace{0cm}\langle D^{*0}(\epsilon,p) \pi^0(q) | \tilde{O}_1(0)
|\bar{B}^0(p+q) \rangle &=&-\frac{\epsilon^*\cdot q~ m_b~
m_{D^{*}}}{4\sqrt{2}\pi^2 f_{D^*} f_B m_B^2(m_B^2 - m_{D^*}^2)}
\int_{m_c^2}^{s_0^{D^*}} ds~ e^{(m_{D^*}^2-s)/T}\nonumber\\
&&\hspace{-5.5cm}\times\int_{u_0^B}^1 \frac{du}{u} e^{(m_B^2-(m_b^2
- m_{D^*}^2(1-u))/u)/T'} \int_{\frac{m_c^2}{s}}^{1}dt\Bigg
\{f_{3\pi}\Bigg[\int_0^u
\frac{dv}{v^2} \phi_{3\pi}(1 -u, u-v, v)\nonumber\\
&&\hspace{-5.5cm}\times \left(\frac{m_b^2 - m_{D^*}^2}{u} \left[2vt
+(1-3t)\chi\right]+ \left(s -
\frac{m_c^2}{t}\right)(1-t)\right)\nonumber\\
&& \hspace{-5.5cm}+(t-1)\left( s-\frac{m_c^2}{t} \right) \left[
\frac{1}{v^2} \phi_{3\pi}(1 -u, u-v, v) \right ]_{v=0}
\chi\nonumber\\
&& \hspace {-5.5cm}+\left(2(2t-1)\frac{m_b^2 - m_{D^*}^2}{u}
\frac{1}{v^2} \phi_{3\pi}(1 -u, u-v, v)
\right .\nonumber\\
&&\hspace{-5.5cm}\left . +(t-1)\left( s -
\frac{m_c^2}{t}\right)\frac{\partial}{\partial v
}\left[\frac{1}{v^2} \phi_{3\pi}(1 -u, u-v, v)\right]\right
)_{v=0}\chi^2\Bigg]\nonumber\\
&& \hspace{-5.5cm}+ m_b f_{\pi}\Bigg[\int_0^u
\frac{dv}{v^2}\Bigg(3v\left[(2t-1){\phi}_{\perp}(1 -u, u-v, v)+
\tilde{\phi}_{\perp}(1 -u, u-v, v)\right]\nonumber\\
&&\hspace{-5.5cm}- 2\left[2(2t-1){\phi}_{\perp}(1 -u, u-v, v)+
\tilde{\phi}_{\perp}(1 -u, u-v, v)\right]\chi\Bigg)\nonumber\\
&&\hspace{-5.5cm}+\frac{1}{2}\left(\frac{3}{v^2}\left[(2t-1){\phi}_{\perp}(1-u,
u-v, v)+\tilde{\phi}_{\perp}(1-u, u-v, v)\right]\right .\nonumber\\
&&\hspace{-5.5cm}\left .+\frac{1}{v}\frac{\partial}{\partial v
}\left[(2t-1){\phi}_{\perp}(1-u, u-v,v)-\tilde{\phi}_{\perp}(1-u,
u-v, v)\right]\right)_ {v=0}\chi^2\Bigg]\Bigg \},
\label{eq:finalresult}
\end{eqnarray}
where $u_0^B = (m_b^2 - m_{D^*}^2)/(s_0^B - m_{D^*}^2)$ with $s_0^B$
being the QCD effective threshold in $B$ channel, $T'$ indicates the
Borel parameter corresponding to the variable $(p+q)^2$, the pionic
distribution amplitude $\phi_{\perp}~(\tilde{\phi}_{\perp})$ is of
twist-4 \cite{LCF}.

To go on with numerical discussions. We take \cite{Input1, Input2}
$m_D^{*0}=2.007$ GeV, $m_c=1.3\pm 0.1$ GeV, $f_{D^*}=240\pm 10$ MeV
$, s_0^D=6\pm 1$ GeV$^2$ and $T=1-2$ GeV$^2$ for the $D^{*0}$
channel parameters, and  $m_B=5.28$ GeV, $m_b=4.7\pm 0.1$ GeV,
$f_B=180\pm 30$ GeV, $s_0^B=35\pm 2$ GeV$^2$ and $T'=8-12$ GeV$^2$
for the $B$ channel ones. The nonperturbative quantities concerning
$\pi$ meson are chosen as \cite{Input2}: $f_{\pi}=132~MeV$, $f_{3
\pi}(\mu_b) = 0.0026~GeV $, and
\begin{eqnarray}
&&\varphi_{3\pi}(\alpha_i,\mu_b)=360\alpha_1 \alpha_2 \alpha_3^2\left[1+\frac{\omega_{3\pi}}{2}(7\alpha_3-3)\right],\nonumber \\
&&\varphi_{\perp}(\alpha_i,\mu_b)=30\delta_{\pi}^2(\alpha_1-\alpha_2)\alpha_3^2[\frac{1}{3}+2\epsilon_{\pi}(1-2\alpha_3)],\nonumber \\
&&\tilde{\varphi}_{\perp}(\alpha_i,\mu_b)=30\delta_{\pi}^2(1-\alpha_3)\alpha_3^2\left[\frac{1}{3}+2\epsilon_{\pi}(1-2\alpha_3)\right],
\end {eqnarray}
whth $\mu_b = \sqrt{m_B^2 - m_b^2}\sim m_b/2$, a scale
characteristic of the typical virtuality of the underlying $b$
quark, and $\omega_{3\pi}=-0.28$, $\delta^2_{\pi}=0.17
\textsl{GeV\,}^2$, $\epsilon_{\pi}=0.36$. The other parameters are
fixed as $C_1(\mu_b)=-0.26$, $C_2(\mu_b)=1.12$ and
$F_1^{B\to\pi}(m^2_{D^*})=0.30$ \cite{Input1}.

Using these inputs, we see that with both the Borel parameters
varying in the respective intervals, Eq.(\ref{eq:finalresult})
reveals a good stability in numerical value. From the sum rule
$\emph{windows}$, the resulting impact on the decay amplitude could
be derived using
\begin{equation}
{\cal A}_{S}(\bar{B}^0\to D^{*0}\pi^0)=\sqrt{2}G_F V_{cb}V^*_{ud}C_2
\langle D^{*0}(\epsilon, p) \pi^0(q) | \tilde{O}_1(0)
|\bar{B}^0(p+q) \rangle.
\end{equation}
Nevertheless, for the purpose of comparing with the factorizable
amplitude and reducing the uncertainty, it is more advisable to
calculate the ratio of ${\cal A}_{S}$ over the factorizable
amplitude ${\cal A}_{F}$. The result is
\begin{eqnarray}
R&=&{\cal A}_S(\bar{B}^0\to D^{*0}\pi^0)/{\cal
A}_F(\bar{B}^0\to D^{*0}\pi^0).\nonumber \\
&=&0.43-0.95.
\end{eqnarray}
The range quoted here is due to the variations of the various inputs
and is obtained by adding linearly up the corresponding
uncertainties. The soft contribution is shown to be of the same
phase as the factorizable one, indicating that it does not provide
the decay amplitude with any strong phase, and with a certain
uncertainty such effect modifies the magnitude of the factorizable
amplitude at a comparable or even identical level.

When a comparison is drawn with the situation of $\bar{B}^0\to
D^{0}\pi^0$ in which the ratio is estimated at between 0.54 and 1.15
\cite{Cui}, the decay in question receives a little less soft
modification within the framework of LCSR, as expected. Also, It is
easily figured out by intuitive picture that a bit more ratio is
observed in the above two cases than $R=0.30-0.70$ yielded for $B\to
J/\psi K $ in \cite{Melic}.

To more give prominence to the role of the soft effects, we might,
as in Eq(3), express ${\cal A}={\cal A}_F+{\cal A}_S$ as
\begin{equation}
{\cal A}(\bar{B}^0\to D^{*0}\pi^0)=-\frac{1}{2}G_F
V_{cb}V^*_{ud}(m_b^2-m_{D^*}^2)f_{D^*}
F_1^{B\to\pi}(m^2_D)a_2^{eff},
\end{equation}
in term of a phenomenological parameter $a_2^{eff}=
a_2^{f}+2C_2R_2$, with $a_2^f= C_1+C_2/3$ corresponding to the
factorizable contribution and $R_2$ being introduced to parameterize
the soft effect. While $R_2$ turns out numerical smallness, the
resulting influence on $a_2^{eff}$ is considerable, due to the large
coefficient $2C_2$ multiplying it. Numerically, the soft effect is
such that it adds a positive number to $a_2^f$, changing the value
of $a_2^{eff}$ from 0.12 to $0.16- 0.23$. Such a result, along with
that obtained in $\bar{B}^0\to D^0\pi^0$, keeps still away from that
demanded by the experiment data \cite{data} $\mid a_2({\cal
A}(\bar{B}^0\to D^{0(*0)}\pi^0)\mid=0.57\pm 0.06$.

At this point, a minor comments should be in order on a couple of
open problems with soft dynamics in charmed color suppressed B
decays: As a result of the dominance of nonfactorizable soft
dynamics in $\bar{B}^0\to D^{0(*0)}\pi^0$, in addition to transverse
momentum dependence which is important to get a result of infrared
finiteness, the higher-twist components have to be counted in the
light cone wavefunction description of $D^{0(*0)}$ meson. The same
must also be done for $B\to J/\psi K$, notwithstanding the fact that
no corresponding infra divergence appears so as to validate QCD
factorization for it, owing to the emitted quarkonium system $(C\bar
C)$ being of a small size $\sim 1/\alpha_s(m_c)m_c\ll
1/\Lambda_{QCD}$. On the other hand, as shown in $B\to \pi\pi,
J/\psi K $, the hard spectator contributions have a divergent
integral from the end point region of the distribution amplitude for
the light mesons, when the twist-3 components is included. However,
the transverse momenta may not be omitted in the region. We believe
that for $\bar{B}^0\to D^{0(*0)}\pi^0 $ it is this case too. The
other resources of soft effects, of course, such as final state
interactions \cite{FSI}, have to enter consideration in an effective
manner for understanding the data.

Colligating the LCSR analyses made for $\bar{B}^0\to D^{0(*0)}\pi^0,
J/\psi~K$, we conclude that nonfactorizable higher twist
corrections, in spite of being a nonleading effect, are important to
understand the charmed color suppressed B decays. These findings
favor, in particular, the argument for QCD factorization:
lange-distance QCD dynamics predominates in the color-suppressed
processes such as $\bar{B}^0\to D^{0(*0)}(\pi^0,\eta,\eta')$. 
A further investigation is necessary to improve the present results.
The staple improvement is to take into account the QCD radiative
correction to the correlator. The other mends are going to wait
until the higher-twist amplitudes of pion become updated and
available.

\newpage
\begin{figure}[htbp] \vspace{0.0cm}
\begin{center}
\includegraphics[scale=0.8]{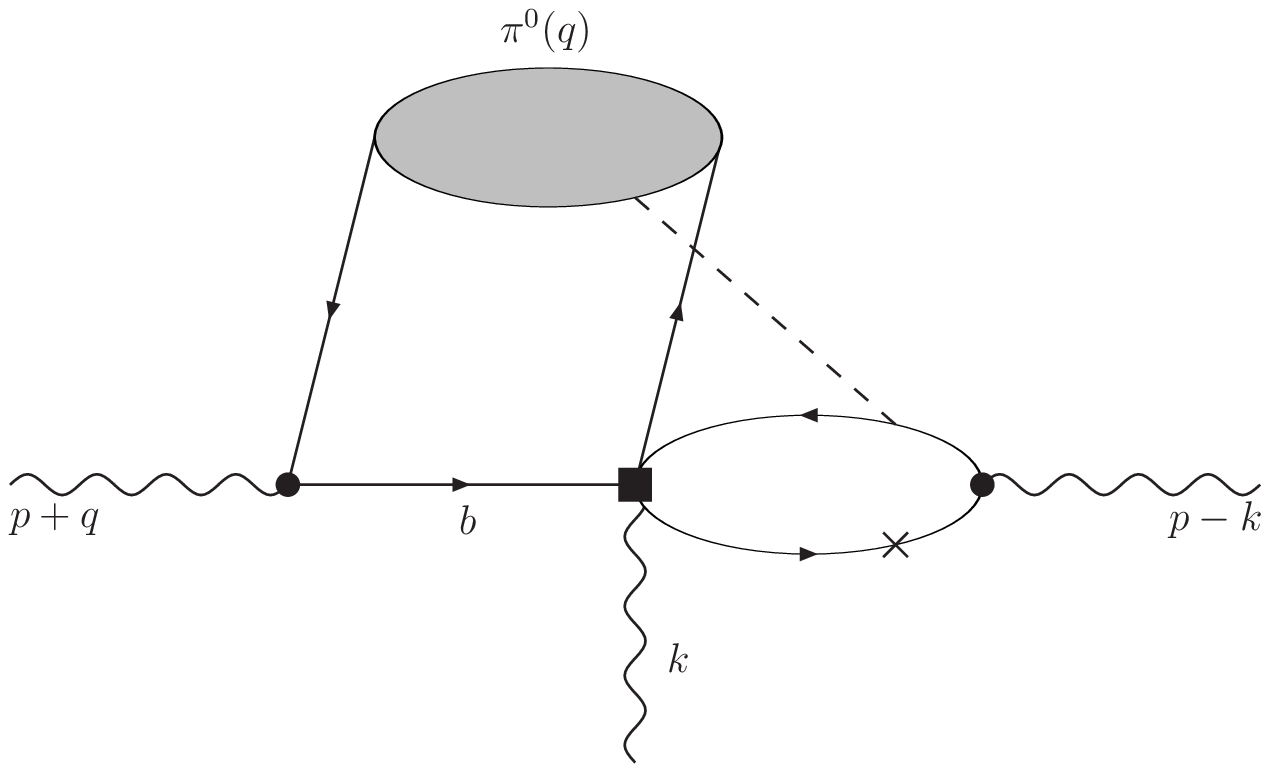}
\vspace{-13.cm} \caption{Feynman diagram illustrating
nonfactorizable higher twist contributions to the correlator (6).
Solid lines represent quarks, dashed line gluon, wavy lines stand
for external currents. The square refers to the insertion of the
operator $\tilde{{\cal O}_1}$, and oval the pionic higher twist
wavefunctions. The cross denotes the other point to which gluon line
can be attached.}
\end{center}
\end{figure}

\end{document}